\documentclass[12pt]{article}
\usepackage{graphicx}

\newcommand\pubnumber{CIPANP2015-Kanazawa}
\newcommand\pubdate{\today}

\def\philly{Department of Physics, SERC, Temple University,
Philadelphia, Pennsylvania 19122, USA}

\def\niigata{Department of Physics, Niigata University, Ikarashi,
Niigata 950-2181, JAPAN}

\def\upton{RIKEN BNL Research Center, Brookhaven National Lab, Upton,
New York 11973, USA}

\def\support{\footnote{Work supported by the National Science
Foundation under Contract No. PHY-1205942.}}

\def\supportyk{\footnote{Work supported by the Grant-in-Aid for
Scientific Research from the Japanese Society of Promotion of Science
under Contract No. 26287040.}}

\def\supportdp{\footnote{Work supported by the RIKEN BNL Research Center.}}

\def\Title#1{\begin{center} {\Large #1 } \end{center}}
\def\Author#1{\begin{center}{ \sc #1} \end{center}}
\def\Address#1{\begin{center}{ \it #1} \end{center}}

\newcommand\pubblock{\rightline{\begin{tabular}{l} \pubnumber\\
         \pubdate  \end{tabular}}}
\newenvironment{Abstract}{\begin{quotation}  }{\end{quotation}}
\newenvironment{Presented}{\begin{quotation} \begin{center} 
             PRESENTED AT\end{center}\bigskip 
      \begin{center}\begin{large}}{\end{large}\end{center} \end{quotation}}

\def\beq{\begin{equation}}
\def\eeq#1{\label{#1}\end{equation}}
\def\eeqn{\end{equation}}

\def\beqa{\begin{eqnarray}}
\def\eeqa#1{\label{#1}\end{eqnarray}}
\def\eeqan{\end{eqnarray}}

\let\bar=\overbar

\def\Dslash{\not{\hbox{\kern-4pt $D$}}}
\def\dslash{\not{\hbox{\kern-2pt $\del$}}}

\def\msb{{\bar{\ssstyle M \kern -1pt S}}}

\newcommand{\vb}{\biggl|}
\newcommand{\pup}{p^{\uparrow}}

\begin{document}
\begin{titlepage}
\pubblock

\vfill
\Title{Collinear Twist-3 Approach to Transverse Single-Spin Asymmetry in
 Proton-Proton Collision}
\vfill
\Author{ Koichi Kanazawa$^1$ and Andreas Metz\support}
\Address{\philly}

\Author{Yuji Koike\supportyk}
\Address{\niigata}

\Author{Daniel Pitonyak\supportdp}
\Address{\upton}

\vfill
\begin{Abstract}
We present our recent analysis on the transverse single-spin asymmetry
(SSA) in inclusive pion and direct-photon production in $pp$ collisions for
RHIC kinematics. The analysis includes the contributions from twist-3
 quark-gluon-quark correlations in the 
 proton and
 twist-3 fragmentation effects for the pion. Some of the functions
 appearing in the formula for the SSA, such as the soft-gluon-pole
 Qiu-Sterman function, the nucleon transversity, and the Collins
 function were fixed consistently with the SSA data in semi-inclusive
 DIS and in $e^+e^-$-annihilation, so that our analysis is free from the
 sign mismatch problem.
\end{Abstract}
\vfill
\begin{Presented}
Conference on the Intersections of
Particle and Nuclear Physics (CIPANP)\\
Vail, CO USA May 19--24, 2015
\end{Presented}
\vfill
\end{titlepage}
\def\thefootnote{\fnsymbol{footnote}}
\setcounter{footnote}{0}

\section{Introduction}

Clarification of the origin of transverse single-spin asymmetry
(SSA) $A_N$ has remained an important issue in high energy hadron
physics since early measurements of $A_N^\pi$ at E704 reported striking
results that the asymmetries are up to 30$\%$ at
forward rapidity~\cite{Adams:1991rw,Adams:1991cs}. Subsequently, this novel spin
phenomenon has also been observed at higher energies at
RHIC~\cite{Adams:2003fx,:2008qb,Lee:2007zzh,:2008mi,Adamczyk:2012xd,Adare:2013ekj}.
Since such large $A_N$ cannot be explained within the collinear parton
model~\cite{Kane:1978nd}, it requires an extension of the framework for
QCD hard processes.

%
One such extension is the TMD factorization approach, which is
applicable for low-$P_T$ processes which contain a separate hard scale 
$Q$ ($\gg P_T$) for a perturbative treatment. 
%
%
%
There, 
large $A_N$ is attributed to the nonzero Sivers/Collins
function which embodies the correlation between parton's intrinsic
transverse momentum and hadron/quark spin. These functions have been
extracted through
phenomenological analyses of data in SIDIS and
$e^+e^-$~\cite{Anselmino:2008sga,Anselmino:2013rya,Anselmino:2013vqa}.
On the other hand, in high-$P_T$ reactions, where $P_T$ is regarded as a
hard scale, large $A_N$ can be
described by means of twist-3 multiparton correlation effects in the
framework of the collinear factorization. 
Of particular interest in the collinear twist-3 approach was
 the chiral-even soft-gluon-pole (SGP) function, also known as the
Qiu-Sterman (QS) function. This effect was believed to be the main
 source of $A_N^\pi$ for many
years~\cite{Qiu:1991pp,Qiu:1991wg,Qiu:1998ia}, and in fact the 
 inclusion of this effect could lead to a reasonable description of RHIC
data~\cite{Kouvaris:2006zy,Kanazawa:2010au,Kanazawa:2011bg}. 
%
%
%
Recently, however, a challenging observation regarding consistency
between the two approaches was made in Ref.~\cite{Kang:2011hk}, where it
was argued that the QS function determined from $pp$ data has an
opposite sign to the one expected from SIDIS data. This
infamous ``sign-mismatch'' problem cannot
be resolved with a more flexible parameterization of the Sivers function
$pp$~\cite{Kang:2012xf}, so that it has been recognized that 
the QS effect cannot be the main source of $A_N^\pi$
in $pp$. 
Another
evidence of this statement is the fact that within the collinear twist-3
approach one obtains the wrong sign for the neutron $A_N$ in inclusive
DIS when using the QS function extracted directly from $A_N^\pi$ in
$pp$~\cite{Metz:2012ui}. Given these new progresses, it has become
important to figure out what is the main cause of $A_N^\pi$ as well
as to test current knowledge on the QS function by getting additional
information from the SSA in other processes.

In this report, we address these issues by looking into $A_N^\pi$ and
$A_N^\gamma$ in $pp$. In the
first part, we present our latest analysis of RHIC data on $A_N^\pi$. We
demonstrate that the twist-3 fragmentation function can play a central
role in the description of $A_N^\pi$ and including this contribution 
leads to a unified description of the asymmetries in $pp$, SIDIS, and
$e^+e^-$. 
In the latter part, we show our estimate of $A_N^\gamma$ at 
RHIC kinematics based on the polarized cross section for the
twist-3 quark-gluon correlation which recently we completed. 
Making a comparison with the one from the TMD approach, we argue that
measurements of $A_N^\gamma$ can help discriminate between these
approaches. This report is a short summary of our recent
papers~\cite{Kanazawa:2014dca,Kanazawa:2014nea}.

\section{Fitting of $A_N^\pi$ at RHIC}

In principle 
$A_N^\pi$ in $pp$ receives contribution from both the twist-3
distribution and fragmentation functions. 
The former piece has been extensively studied in the literature 
\cite{Qiu:1998ia,Kouvaris:2006zy,Koike:2009ge} while the
complete cross section formula for the latter has become available very
recently~\cite{Kang:2010zzb,Metz:2012ct}. Interestingly, a
phenomenological study with a simple model in Ref.~\cite{Kang:2010zzb}
shows that the latter effect could be significant at RHIC
kinematics.
Here we perform a new fit of RHIC $A_N^\pi$
data~\cite{Adams:2003fx,:2008qb,Adamczyk:2012xd,Lee:2007zzh}
by including the whole contribution from the twist-3 fragmentation
function and address whether this can resolve the sign-mismatch problem. 

Let us begin with the setting for our fits. 
In order to evade the sign-mismatch problem on the QS function $T_F(x,x)$, 
we use the existing parameterization of the Sivers function
$f_{1T}^\perp$ which was extracted from SIDIS data.
%
%
The rigorous relation between these functions
reads~\cite{Boer:2003cm}
\begin{equation}
T_F(x,x) = - \int d^2\vec{p}_\perp \frac{\vec{p}_\perp^{\,2}}{M}
 f_{1T}^\perp(x,\vec{p}_{\perp}^{\,2}) \vb_{\rm SIDIS},
\label{e:siv}
\end{equation}
where the Sivers function is the one which shows up in SIDIS. In our
analysis we try two different parametrizations in
Ref.~\cite{Anselmino:2008sga,Anselmino:2013rya}.
Similarly, for the fragmentation contribution, we take advantage of the
existing parameterization of the Collins function as well as take into
account the relation between the relevant twist-3 functions in order to
make our analysis consistent.
%
First, we note the cross section formula for fragmentation contains
three twist-3 fragmentation functions: $H$, $\hat{H}$, and
$\hat{H}_{FU}^\Im$. Of these
three, by invoking the EOM relation~\cite{Metz:2012ct}
\begin{equation}
 H(z) = - 2 z \hat{H}(z) 
  + 2 z^3 \int_{z}^{\infty} \frac{dz_1} {z_1^2} \frac{1}
  {\frac{1}{z} - \frac{1}{z_1}} \hat{H}_{FU}^\Im(z,z_1),
\label{e:eom}
\end{equation}
one can choose $\{\hat{H}, \hat{H}_{FU}^\Im\}$ as independent functions.
In this way $H(z)$ is regarded an auxiliary function and is completely
determined by the other two.
The kinematical twist-3 fragmentation function $\hat{H}(z)$ can
be fixed in terms of the TMD Collins function via the relation
\begin{equation}
 \hat{H}(z) = z^2 \int d^2 \vec{k}_\perp \frac{\vec{k}_\perp^2}{2
  M_h^2} H_{1}^\perp(z,z^2\vec{k}_\perp^2).
 \label{e:Hhat}
\end{equation}
For the Collins function $H_1^\perp$, we take the
parameterization 
from Ref.~\cite{Anselmino:2013vqa}. 
Concerning the 3-parton correlator $\hat{H}_{FU}^\Im$, so far no knowledge has
 been obtained as it has no counterpart in the TMD approach, and
 therefore this function needs to be determined by fitting $pp$
data. We refer the reader to Ref.~\cite{Kanazawa:2014dca} for our
parameterization as well as any other details of our fits. 

\begin{figure}[htb]
\centering
\includegraphics[height=1.15in]{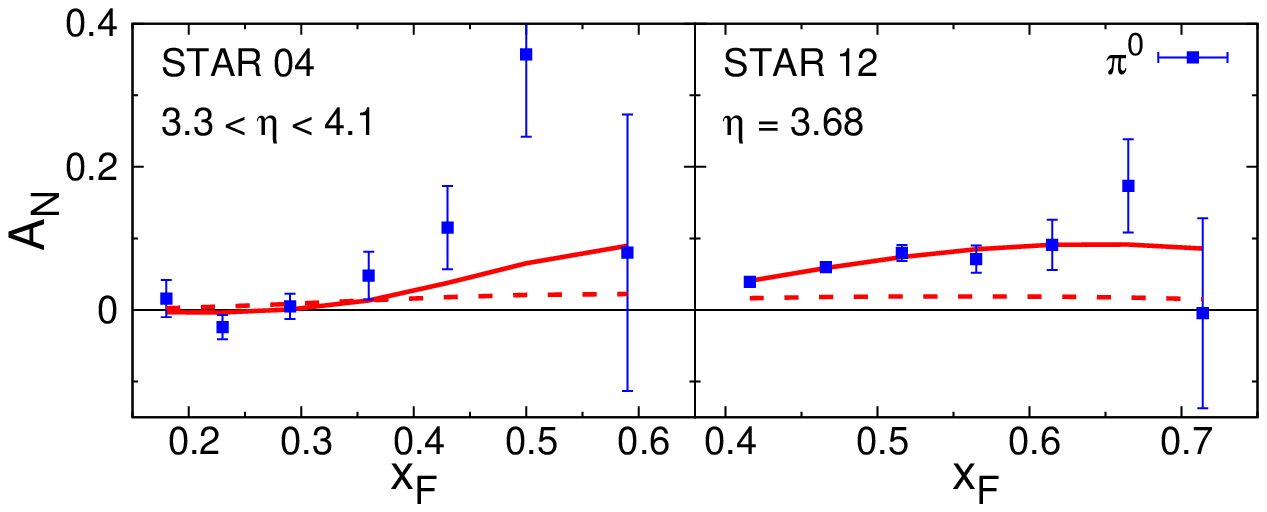}
\includegraphics[height=1.15in]{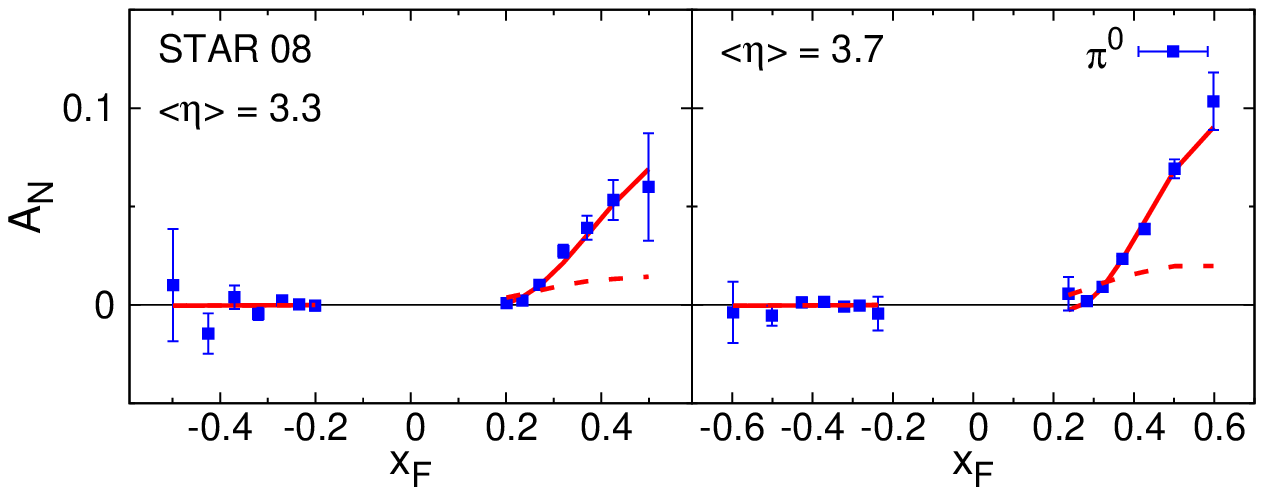}
\includegraphics[height=1.15in]{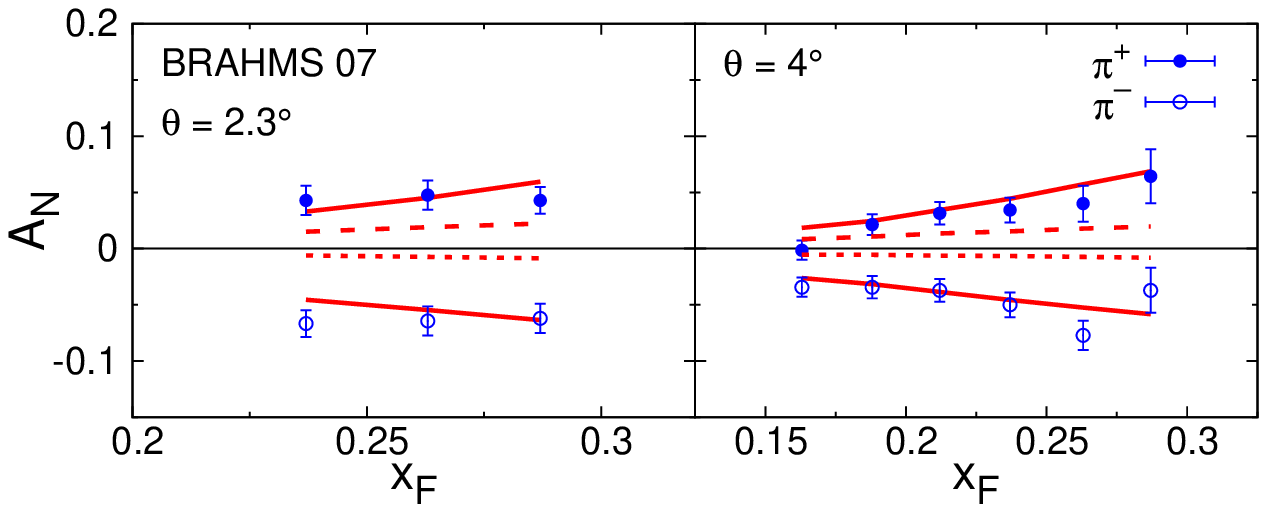}
\caption{Fit result for pions. We included only the data at $\sqrt{S}=200$
 GeV from STAR and
 BRAHMS~\cite{Adams:2003fx,:2008qb,Adamczyk:2012xd,Lee:2007zzh}. The
 dashed curves are calculated without including $\hat{H}_{FU}^\Im$.}
\label{fig:fit}
\end{figure}

\begin{figure}[htb]
\centering
\includegraphics[height=3in]{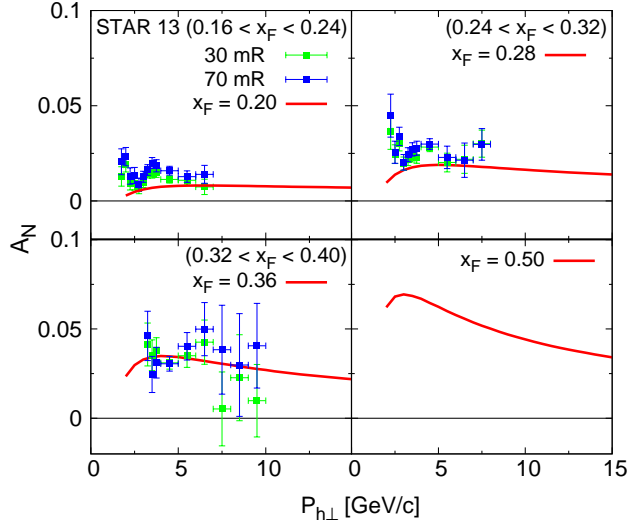}
\caption{Comparison of $P_T$-dependence of $A_N^\pi$ with the
 STAR data~\cite{Heppelmann:2013ewa}.}
 \label{fig:pt}
\end{figure}

Figure \ref{fig:fit} shows the result of our
fits. Overall, the $pp$ data are successfully reproduced
both for neutral and charged pions as a function of $x_F$, and we found
most of the contribution come from the twist-3 fragmentation function.
Also, we plotted the dashed curves which show the calculation without
including the contribution from the 3-parton correlator, namely it
represents the contribution of $\hat{H}$.
Obviously, this contribution is insufficient
to reproduce the RHIC data and we found indeed the 3-parton correlator
gives the dominant contribution and thus plays a crucial role in the
description of the RHIC data. 

In Fig.~\ref{fig:pt}, we have presented our prediction for the
$P_T$-dependence of $A_N$ which is compared to the latest STAR data
in~\cite{Heppelmann:2013ewa}.
One sees that the observed pattern is reproduced as well, supporting the
validity of the collinear twist-3 approach.


\section{Phenomenology of $A_N^\gamma$ at RHIC}

\begin{figure}[htb]
\centering
\includegraphics[height=1.3in]{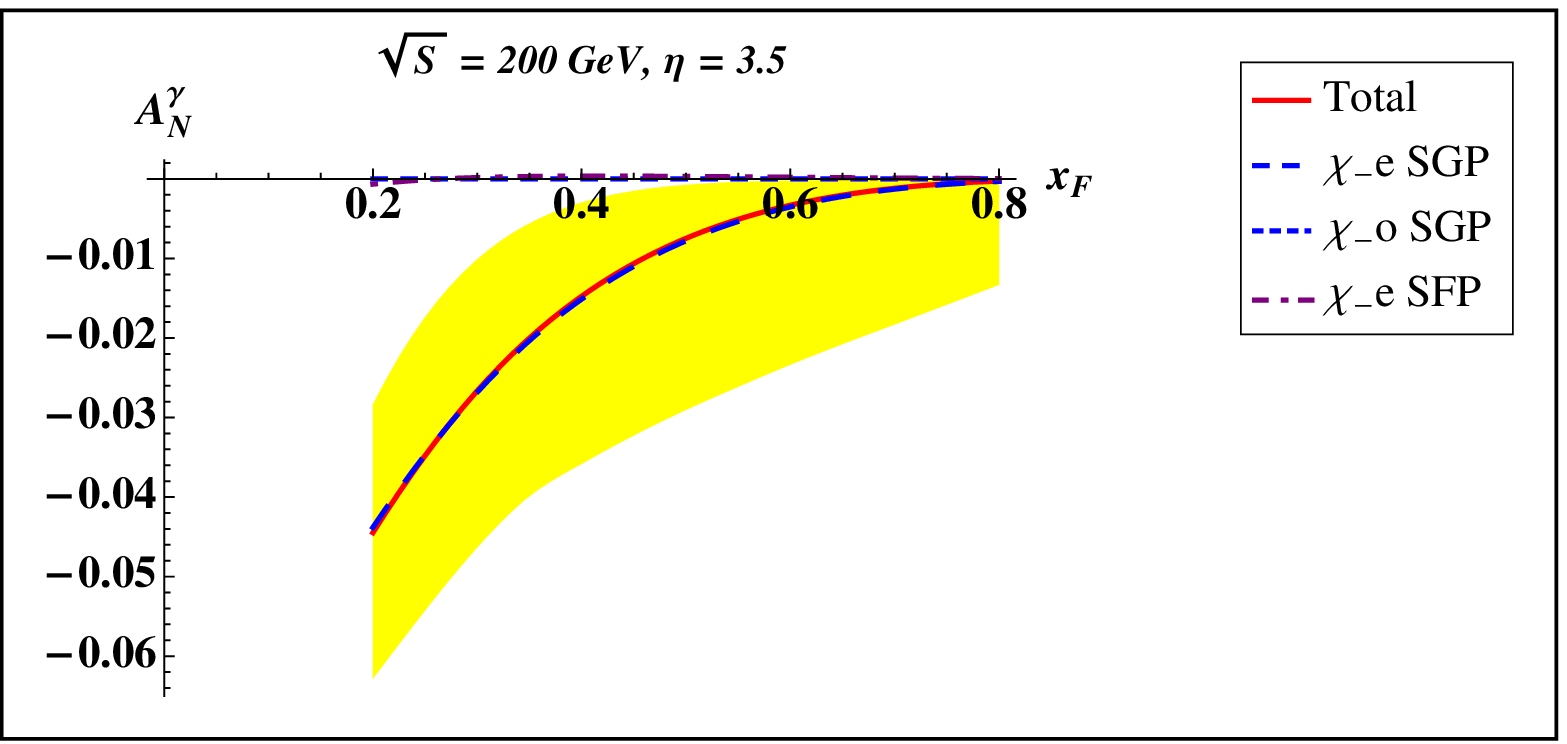}
\includegraphics[height=1.3in]{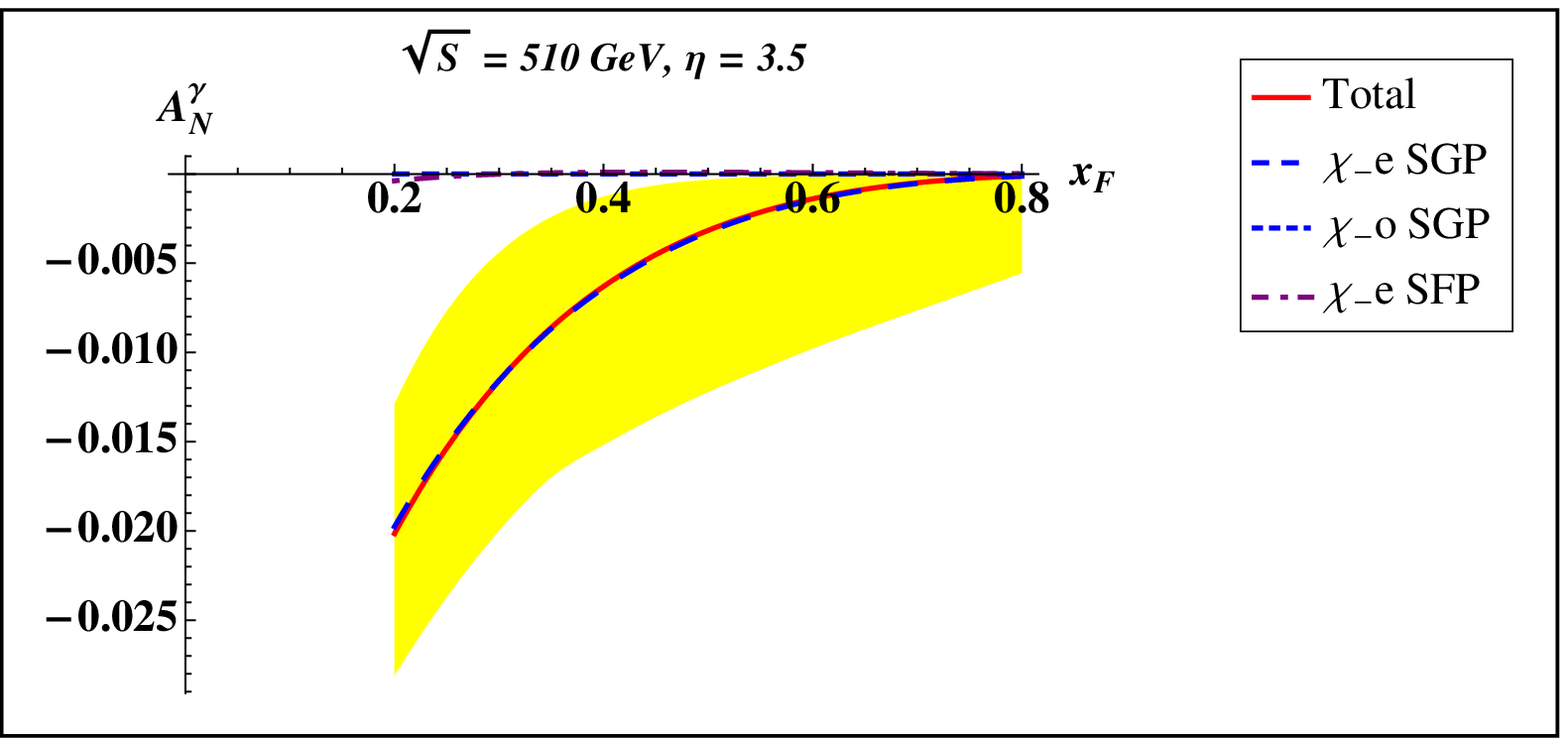}
\caption{Estimate of $A_N^\gamma$ at forward pseudorapidity $\eta=3.5$
 at two different energies.}
\label{fig:photon}
\end{figure}

The SSA in direct-photon production in $pp$ is a clean process to probe
the twist-3 distributions inside proton because of the absence of
fragmentation. 
By now the cross section for the quark-gluon and 3-gluon
correlations in $\pup$ is available in
Ref.~\cite{Qiu:1991pp,Qiu:1991wg,Ji:2006vf,Kanazawa:2011er,Koike:2011nx},
where it was shown that the former effect to $A_N^\gamma$ is much larger
than the case of $A_N^{\pi^0}$ at forward rapidity  while the latter becomes
significant only at backward rapidity.
In addition, in Ref.~\cite{Kanazawa:2014nea} we have derived the cross
section for the chiral-odd quark-gluon correlation in $p$. 
With the complete formula in hand, here we give a new numerical estimate
of  forward $A_N^\gamma$ at RHIC kinematics and see what we can learn
from it.

The contribution from the quark-gluon correlation is classified into the
ones from SGP and soft-fermion-pole (SFP), so in total there are four types of
contributions: (i) chiral-even SGP (QS effect), (ii) chiral-even SFP,
(iii) chiral-odd SGP, and (iv) chiral-odd SFP. Of these four, in
Ref.~\cite{Kanazawa:2014nea} we found there is no contribution from (iv)
at leading-order QCD, so we focus on the other three.
As in the case of $A_N^\pi$, we perform our calculation of $A_N^\gamma$
consistently with the existing parametrizations of the relevant TMD
functions. For this purpose, we again make use of Eq.~(\ref{e:siv}) to
fix the QS function. 
Likewise, we use another rigorous relation between the chiral-odd SGP
function $E_F(x,x)$ and the Boer-Mulders function
$h_1^\perp$~\cite{Boer:2003cm}
\begin{eqnarray}
 E_F(x,x) = \pm \frac{1}{\pi M^2} \int d^2 \vec{p}_\perp \,
  \vec{p}_\perp^{\, 2}
  h_1^{\perp (\pm)} (x, \vec{p}_\perp),
\end{eqnarray}
where the sign $+ (-)$ represents that the future-pointing
(past-pointing) Wilson line is used in the definition of the
Boer-Mulders function.
For the Boer-Mulders function we take the parameterization from
Ref.~\cite{Barone:2010gk}.
The only unknown input is the chiral-even SFP function
$T_F(0,x)+\tilde{T}_F(0,x)$. For this function, we make a simple
assumption as
\begin{eqnarray}
 T_F(0,x) + \tilde{T}_F(0,x) &=& T_F(x,x),
\end{eqnarray} 
to see its impact at the RHIC energy. 

Figure~\ref{fig:photon} shows our estimates of forward $A_N^\gamma$ at
$\sqrt{S}=200$ GeV and 510 GeV, respectively. We have found at both
energies the
asymmetry could be substantial and has negative sign. Also shown in the
figure is the decomposition into the contributions from each
function. Clearly, the asymmetry is dominated
by the QS effect and the contribution from other
sources are negligible, suggesting $A_N^\gamma$ is an ideal
observable to extract the QS function. 
Another interesting finding is that our result differs in sign from
the prediction based on the TMD approach in
Ref.~\cite{Anselmino:2013rya}. This indicates ongoing and future
measurements of $A_N^\gamma$ are quite useful
to discriminate between the two approaches.  



\section{Summary and outlook}

We have presented our recent analyses on $A_N^\pi$ and
$A_N^\gamma$ at RHIC kinematics. We have demonstrated 
the twist-3 fragmentation function can give the dominant contribution to
$A_N^\pi$ and including this effect leads to a good description of
the RHIC data for neutral
and charged pions. By construction, this analysis is consistent with the Sivers
and Collins mechanisms in the TMD approach, and thus we have attained a
first unified description of the asymmetries in $pp$, SIDIS and
$e^+e^-$. 
%
In addition, we have provided a new prediction on $A_N^\gamma$ based on
the complete cross section for the twist-3 quark-gluon correlation. It
turned out that the QS function is the only possible source to cause a
substantial asymmetry in this process. Interestingly, our result
differs in sign from the prediction based on the TMD approach. We
expect ongoing and future measurements of $A_N^\gamma$ will help
discriminate the two approaches. 



\begin{thebibliography}{99}

\bibitem{Adams:1991rw} 
  D.~L.~Adams {\it et al.} [E581 and E704 Collaborations],
  Phys.\ Lett.\ B {\bf 261}, 201 (1991).

\bibitem{Adams:1991cs} 
  D.~L.~Adams {\it et al.} [E704 Collaboration],
  Phys.\ Lett.\ B {\bf 264}, 462 (1991).

\bibitem{Adams:2003fx}
  J.~Adams {\it et al.} [STAR Collaboration],
  Phys.\ Rev.\ Lett.\  {\bf 92}, 171801 (2004).

\bibitem{:2008qb}
  B.~I.~Abelev {\it et al.} [STAR Collaboration],
  Phys.\ Rev.\ Lett.\  {\bf 101}, 222001 (2008).
  
\bibitem{Adamczyk:2012xd} 
  L.~Adamczyk {\it et al.}  [STAR Collaboration],
  Phys.\ Rev.\ D {\bf 86}, 051101 (2012).

\bibitem{Lee:2007zzh} 
  J.~H.~Lee {\it et al.}  [BRAHMS Collaboration],
  AIP Conf.\ Proc.\  {\bf 915}, 533 (2007).

\bibitem{:2008mi} 
  I.~Arsene {\it et al.}  [BRAHMS Collaboration],
  Phys.\ Rev.\ Lett.\  {\bf 101}, 042001 (2008).
    
\bibitem{Adare:2013ekj} 
  A.~Adare {\it et al.} [PHENIX Collaboration],
  Phys.\ Rev.\ D {\bf 90}, 012006 (2014).

\bibitem{Kane:1978nd} 
  G.~L.~Kane, J.~Pumplin and W.~Repko,
  Phys.\ Rev.\ Lett.\  {\bf 41}, 1689 (1978).

\bibitem{Anselmino:2008sga} 
  M.~Anselmino {\it et al.},
  Eur.\ Phys.\ J.\ A {\bf 39}, 89 (2009).

\bibitem{Anselmino:2013rya} 
  M.~Anselmino {\it et al.},
  Phys.\ Rev.\ D {\bf 88}, 054023 (2013).

\bibitem{Anselmino:2013vqa} 
  M.~Anselmino {\it et al.},
  Phys.\ Rev.\ D {\bf 87}, 094019 (2013).

\bibitem{Qiu:1991pp} 
  J.-w.~Qiu and G.~F.~Sterman,
  Phys.\ Rev.\ Lett.\  {\bf 67}, 2264 (1991).

\bibitem{Qiu:1991wg} 
  J.-w.~Qiu and G.~F.~Sterman,
  Nucl.\ Phys.\ B {\bf 378}, 52 (1992).

\bibitem{Qiu:1998ia} 
  J.-w.~Qiu and G.~F.~Sterman,
  Phys.\ Rev.\ D {\bf 59}, 014004 (1999).

\bibitem{Kouvaris:2006zy}
  C.~Kouvaris {\it et al.},
  Phys.\ Rev.\  D {\bf 74}, 114013 (2006).

\bibitem{Kanazawa:2010au} 
  K.~Kanazawa and Y.~Koike,
  Phys.\ Rev.\ D {\bf 82}, 034009 (2010).

\bibitem{Kanazawa:2011bg} 
  K.~Kanazawa and Y.~Koike,
  Phys.\ Rev.\ D {\bf 83}, 114024 (2011).

\bibitem{Kang:2011hk} 
  Z.-B.~Kang {\it et al.},
  Phys.\ Rev.\ D {\bf 83}, 094001 (2011).

\bibitem{Kang:2012xf} 
  Z.-B.~Kang and A.~Prokudin,
  Phys.\ Rev.\ D {\bf 85}, 074008 (2012).

\bibitem{Metz:2012ui} 
  A.~Metz {\it et al.},
  Phys.\ Rev.\ D {\bf 86}, 094039 (2012).

\bibitem{Kanazawa:2014dca} 
  K.~Kanazawa, Y.~Koike, A.~Metz and D.~Pitonyak,
  Phys.\ Rev.\ D {\bf 89}, 111501(R) (2014).

\bibitem{Kanazawa:2014nea} 
  K.~Kanazawa, Y.~Koike, A.~Metz and D.~Pitonyak,
  Phys.\ Rev.\ D {\bf 91}, no. 1, 014013 (2015).

\bibitem{Koike:2009ge} 
  Y.~Koike and T.~Tomita,
  Phys.\ Lett.\ B {\bf 675}, 181 (2009).

\bibitem{Kang:2010zzb} 
  Z.-B.~Kang, F.~Yuan and J.~Zhou,
  Phys.\ Lett.\ B {\bf 691}, 243 (2010).

\bibitem{Metz:2012ct} 
  A.~Metz and D.~Pitonyak,
  Phys.\ Lett.\ B {\bf 723}, 365 (2013).

\bibitem{Kanazawa:2000hz} 
  Y.~Kanazawa and Y.~Koike,
  Phys.\ Lett.\ B {\bf 478}, 121 (2000).

\bibitem{Kanazawa:2000kp} 
  Y.~Kanazawa and Y.~Koike,
  Phys.\ Lett.\ B {\bf 490}, 99 (2000).

\bibitem{Ji:2006vf} 
  X.~Ji, J.~w.~Qiu, W.~Vogelsang and F.~Yuan,
  Phys.\ Rev.\ D {\bf 73}, 094017 (2006).

\bibitem{Kanazawa:2011er} 
  K.~Kanazawa and Y.~Koike,
  Phys.\ Lett.\ B {\bf 701}, 576 (2011).

\bibitem{Koike:2011nx} 
  Y.~Koike and S.~Yoshida,
  Phys.\ Rev.\ D {\bf 85}, 034030 (2012).


\bibitem{Barone:2010gk} 
  V.~Barone, S.~Melis and A.~Prokudin,
  Phys.\ Rev.\ D {\bf 82}, 114025 (2010).

\bibitem{Boer:2003cm} 
  D.~Boer, P.~J.~Mulders and F.~Pijlman,
  Nucl.\ Phys.\ B {\bf 667}, 201 (2003).

\bibitem{Heppelmann:2013ewa} 
  S.~Heppelmann [STAR Collaboration],
  PoS DIS {\bf 2013}, 240 (2013).



%
%
%
%


\end{thebibliography}
\end{document}